\documentclass[twocolumn]{article}
\usepackage[utf8x]{inputenc}
\usepackage[english]{babel}
\usepackage{authblk}
\usepackage{geometry}
\usepackage{amsmath}
\usepackage{amssymb}
\usepackage{natbib} 
\usepackage{graphicx}
\usepackage{xcolor}
\usepackage{abstract}
\bibpunct{(}{)}{,}{a}{}{,}
\usepackage[colorlinks=true,linkcolor=blue,citecolor=blue]{hyperref}

\advance\topmargin 1cm

\title{Cosmological Observational Tests in the JWST Era.~II: The Tolman Test}

\author[1,2]{V.~V.~Tsymbal}
\author[3,4]{A.~A.~Raikov}
\author[5,6]{N.~Yu.~Lovyagin}

\affil[1]{Institute of Astronomy of RAS, 119017, 48 Pyatnitskaya St., Moscow, Russia}
\affil[2]{Special Astrophysical Observatory of RAS, 369167, Nizhnij Arkhyz, Russia}
\affil[3]{Saint Petersburg Branch of the Special Astrophysical Observatory of RAS, 196140, 65 Pulkovskoe Shosse, Saint Petersburg, Russia}
\affil[4]{Main Astronomical Observatory of RAS, 196140, 65 Pulkovskoe Shosse, Saint Petersburg, Russia}
\affil[5]{Saint Petersburg State University, 199034, Universitetskaya Emb., 7/9, Saint Petersburg, Russia, e-mail: n.lovyagin@spbu.ru}
\affil[6]{Saint Petersburg State Marine Technical University, 190008, Lotsmanskaya St. 3, Saint Petersburg, Russia}

\date{}

\begin{document}

\twocolumn[
\maketitle 
\begin{onecolabstract}
		In this work, we investigate a classical cosmological test --- the dependence of galaxy surface brightness on redshift $z$ (the Tolman test). We analyzed 6 860 galaxies with reliably determined spectroscopic redshifts from the ASTRODEEP-JWST photometric catalogue. We find that (a) the mean surface brightness of galaxies indeed decreases with increasing distance, and (b) the observed trend shows a significant departure from the prediction of the standard cosmological model, which expects the mean surface brightness to decline as $\propto (1+z)^{-4}$. 
		
\vspace*{0.7em}
		\textbf{Keywords:} cosmology: observation, cosmology: expanding, cosmology: early Universe
		
\vspace*{0.7em}

\vspace*{0.5em}

\noindent {\slshape This is a preprint of a manuscript accepted for publication in the \textit{Astrophysical Bulletin}, Vol. 81, No. 3, 2026.

}
\vspace*{1cm}
	\end{onecolabstract}
]

\section{Introduction}

In our previous work \citet{raikov2025cosmological}, we considered the cosmological test ``angular size --- redshift''. The importance of such studies was noted in light of the emergence of new observational data and unresolved problems in the standard cosmological model \citep{di2021realm,kamionkowski2023hubble,melia2024cosmic,munoz2024reionization,aghanim2020planck,dolgov2018mass2}, 
as well as in the context of classical and modern works \citep{geller1972test,sandage1988observational,weinberg2008cosmology,orlov2016cosmological} that consider various cosmological models. Recently, the work \cite{leauthaud2025looking} was also published, describing the current status of unresolved problems in the context of the need to prepare for a possible revision of the $\Lambda$CDM concept.

The present paper is devoted to a more more difficult  test proposed in \citet{tolman1930estimation}. The essence of this test is that on local scales the surface brightness of a radiation source does not change with distance. However, in cosmology this is not the case: the observed surface brightness must depend on the distance to the source, and the form of this dependence differs among various cosmological theories. In the standard cosmological model ($\Lambda$CDM), the surface brightness decreases due to the action of the following factors:

\begin{enumerate}
	\item A decrease in the energy of the detected photons due to redshift, which leads to a decrease in total brightness proportional to $(1+z)^{-1}$
	\item An increase in the angular size of the source, which leads to a decrease in surface brightness proportional to $(1+z)^{-2}$
	\item A time delay of the detected photons due to cosmological time dilation, which leads to a decrease in total brightness proportional to $(1+z)^{-1}$.
\end{enumerate}

Thus, within the framework of $\Lambda$CDM the surface brightness of the detected radiation should decrease proportionally to $(1+z)^{-4}$, whereas, for example, within the framework of the static ``tired light'' theory or when applying the linear Hubble law (the naive Euclidean model), only the first factor affects the decrease of the observed surface brightness.

The practical implementation of the Tolman test is substantially more complicated than that of the ``angular size --- redshift'' test. This is because we cannot measure the bolometric brightness of cosmological sources, and we do not have sources with identical brightness at
different $z$. Moreover, even within a small range of redshifts, the emitted brightness of different galaxies in a specific wavelength interval can differ significantly from one another.

For this reason, the number of works known to us that investigate the Tolman test is relatively small. We first note the long series of works \cite{sandage2001tolman,lubin2001tolman,lubin2001tolman2,lubin2001tolman3,sandage2010tolman2}. In these works it was shown that the decrease in the surface brightness of elliptical galaxies with distance is indeed real. Unfortunately, because the authors had access to observations only at relatively low $z$, performing the full Tolman test on cosmological scales was difficult.

As in our previous work, here we perform an analysis of the Tolman test based on the data from the ASTRODEEP-JWST catalogue \citep{merlin2024astrodeep}, which contains more than 500\,000 galaxies at large (up to $\lesssim 20$) redshifts. For the analysis, we selected 6\,860 galaxies from the catalogue with reliably determined spectroscopic redshifts ($z\lesssim 14$).

\section{Data Used}

The ASTRODEEP-JWST catalogue \citep{merlin2024astrodeep} includes information on $531\,173$ galaxies. In particular, for each object in the catalogue, photometric observations in 16 filters (8 HST filters and 8 JWST filters) in the wavelength range $\approx 4300$--$45000$\AA, the measurement uncertainties, and photometric redshifts determined by four different methods are provided. For objects with known spectroscopic redshifts, their values are also included in the catalogue. The total survey area on the sky is approximately $0.2$ square degrees. The ASTRODEEP-JWST catalogue is homogeneous in the data it presents, as the authors combined the HST and JWST images into a common mosaic system and recalibrated them. Photometric fluxes were measured as the sum of pixel brightnesses contained within the so-called effective (optimal) aperture.

To calculate fluxes, the ASTRODEEP-JWST authors used aperture photometry, selecting the best among 9 apertures ranging from $0.2''$ to $5.30''$. The radii of these apertures for each object are also provided in the catalogue. Thus, the mean surface brightness in each filter is calculated by us as the flux in that filter divided by the aperture area.

Surface brightness can also be defined as the ratio of the flux to the area of a circle or ellipse, calculated using the effective radius (the radius containing 50\% of the detected light) or the major and minor axes of the object's ellipse, 

Factors such as the point spread function (PSF) and the loss of outer regions of fainter (more distant) objects in noise undoubtedly affect the shape of the surface brightness distribution across the disk, but they do not affect the total brightness within the aperture (flux), which we use to calculate the mean surface brightness. Naturally, the value of the optimal aperture decreases at higher $z$, but the detected flux also decreases. In addition, the catalogue authors accounted for the measured PSF when computing fluxes in the filters. Therefore, we cannot expect an artificial decrease in the mean surface brightness due to the chosen method.

As in our previous work, we excluded galaxies with low-quality measurements, images blended with neighbouring objects, objects at the edges of mosaics, as well as sources identified as point-like 
(using the criterion $flags<80$). We used only galaxies with reliably determined spectroscopic redshifts. Although this approach limits the sample size and the maximum accessible $z$ values, it allows us to avoid systematic errors characteristic of photometric redshifts. The sample included only objects with JWST observations. As a result, 6\,860 galaxies with confirmed spectroscopic redshifts were selected.

\begin{figure}
	\includegraphics[width=\hsize]{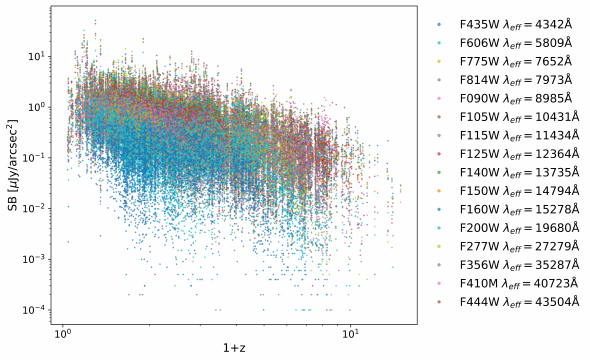}
	\caption{Surface brightness of $6\,860$ galaxies in 16 filters of the ASTRODEEP-JWST catalogue in logarithmic scale. The figure shows that the mean surface brightness decreases with increasing redshift.}
	\label{fig:multifilter}
\end{figure}

Fig. \ref{fig:multifilter} shows the surface brightness data for all galaxies in the sample across all catalogue filters in logarithmic scale. Surface brightness is defined as the ratio of the flux in the filter to the area of the effective aperture. Also shown are the $\lambda_{eff}$ of the filters (data taken from \citeauthor{merlin2024astrodeep}, \citeyear{merlin2024astrodeep}). 

\section{Surface Brightness Determination}

As noted in the introduction, it is practically impossible to select galaxies from the available data that have the same emission intensity in a specific wavelength range and are located at different $z$. This is due not only to observational selection, the diversity of galaxies and their spectra, but also to the fact that for galaxies at different $z$, the same spectral ranges in the galaxy rest frame fall into different detector filters. Therefore, we use several approaches with some physical justification to evaluate the trend of the mean surface brightness as a function of $z$.

The first approach is mean the surface brightness of a galaxy across the entire wavelength range available in the catalogue. Here we approximate the observed quantity as closely as possible to the concept of bolometric brightness used in Tolman's work.

The second approach is to select the maximum observed surface brightness among all 16 observed bands as the characteristic surface brightness of the galaxy. Filters in which the surface brightness is not defined are discarded and not included in the mean value.

\begin{figure}
	\includegraphics[width=\hsize]{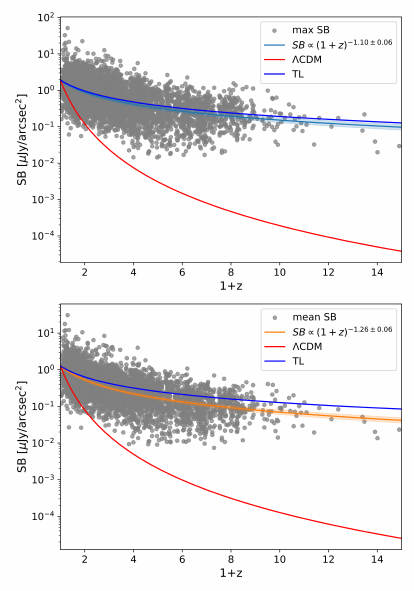}
	\caption{Surface brightness (SB) of $6\,860$ galaxies in comparison with redshift. Top: maximum SB: mean SB across the 16 filters of the ASTRODEEP-JWST catalogue. Observed values are shown as points. Solid lines indicate the best-fit approximation curve with formal error bands, as well as theoretical dependencies for the standard cosmological model $\propto (1+z)^{-4}$ and the ``tired light'' model $\propto (1+z)^{-1}$.}
	\label{fig:SBmaxmean}
\end{figure}

In both cases, for the scatter plots of galaxies in the coordinates $\bigl(1+z, SB\bigr)$, where $SB$ is the catalogue surface brightness in $\mu$Jy/arcsec${}^{-2}$, a simple unweighted least-squares approximation was constructed on a logarithmic scale. The results are shown in Fig. \ref{fig:SBmaxmean}. The plot also shows the theoretical dependencies for the standard cosmological model and the static cosmological model. These are calculated as the change with $(1+z)$ raised to the power of -4 and -1, respectively, from the mean surface brightness of galaxies at $z \to 0$. It can be seen that starting from $z \sim 4$, the observed brightness decrease is significantly smaller than predicted by the standard cosmological model.

The third approach is to consider the dependence of surface brightness on $z$ in the rest frame. For each selected galaxy, the distribution of the mean surface brightness as a function of the effective wavelength of each filter, reduced to the rest frame, is calculated. A common wavelength interval in the rest frame is selected for different $z$, and the mean of the mean surface brightnesses in the filters that fall within this range in the observer frame is computed. Filters in which the surface brightness is not defined are discarded and not included in the mean value.

Due to the fact that the galaxies in the catalogue have a large spread in $z$, a common wavelength interval can be defined only for a limited range of $z$. As already becomes clear from Figs.~\ref{fig:multifilter} and \ref{fig:SBmaxmean}, the effect of surface brightness decrease with distance becomes especially noticeable at large $z$. Moreover, one should expect that the sample of observed galaxies at these $z$ becomes more homogeneous as a result of evolution. For this reason, we selected the range $3 < z < 14$ and the wavelength interval 1500--2000 \AA{} in the rest frame. The resulting sample contains $1\,507$ galaxies. This range corresponds predominantly to the emission of young stars. Although the choice of such a rest-frame interval may be suboptimal, it is dictated solely by the requirement that it must be common for the widest possible range of $z$. Note that the emission intensity of young stars does not depend on $z$ and, therefore, we also achieve the stated goal --- the selection of a quasi-standard emission source.

\begin{figure}
	\includegraphics[width=\hsize]{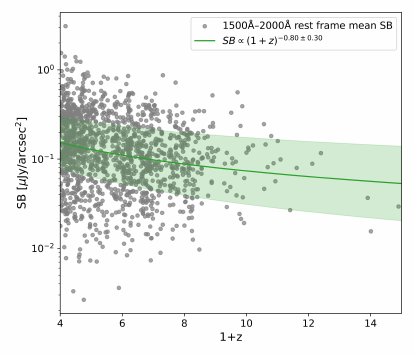}
	\caption{Mean surface brightness in the range 1500--2000 \AA{} for $1\,507$ galaxies compared with redshift. Solid lines show the best-fit approximation curve with bands of formal errors.}
	\label{fig:SBrest}
\end{figure}

The results are shown in Fig. \ref{fig:SBrest}. The diagram shows that the observed decrease in the surface brightness with the distance to the galaxy is somewhat weaker than in Fig. \ref{fig:SBmaxmean}, and the scatter of the values is larger. 

Unfortunately, since this work is devoted to testing cosmological models, we cannot apply a selection based on, for example, absolute magnitudes, such as $M_{UV}$, because such quantities are defined only within the framework of a specific cosmological model.

Instead, in order to partially compensate for the effect of observational selection, we performed the following analysis. The entire range in $z$ is divided into bins. In each bin, a histogram of the surface brightness distribution is constructed. In this histogram, the bin with the maximum number of galaxies $N$ (the mode) is identified. Assuming a Poisson statistics for the distribution of galaxies by surface brightness, bins are selected for which the number of galaxies exceeds $N-\sqrt{N}$. The spread of these histogram bins is taken as an estimate of the error, and the weighted mean (with weights equal to the number of galaxies) of the surface brightness is taken as the characteristic value of the surface brightness at the given $z$ (the centre of the bin in $z$).

The surface brightness selected in this way can, from a physical point of view, be considered the surface brightness of the ``most frequently occurring'' (most typical) galaxy at a given $z$. This approach significantly reduces the influence of observational selection and partially tests the robustness of the method with respect to the diversity of galaxy spectra and their redshifts when determining the surface brightness in a filter in the observer frame.

\begin{figure*}[t]
	\includegraphics[width=\hsize]{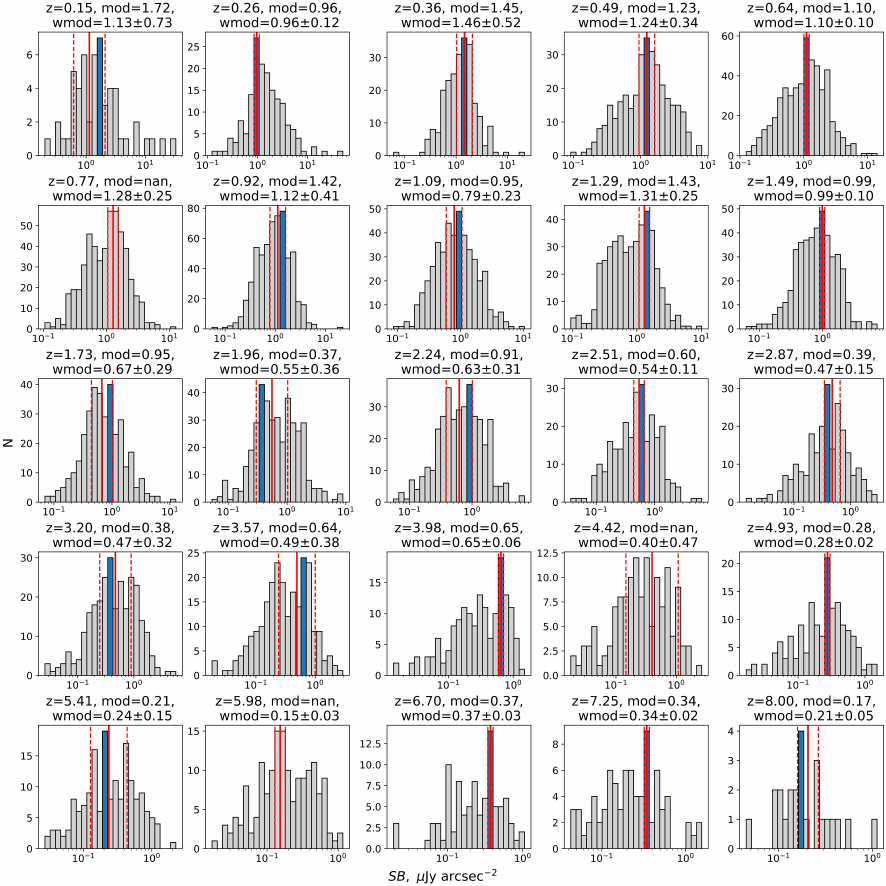}
	\caption{Histograms of surface brightness in bins of $z$. The midpoints of the $z$ bins, the modal surface brightness values (mod, the column highlighted in colour), the weighted modal value (wmod), and the spread of bins close to the modal ones (error of wmod and vertical bars) are shown. The surface brightness of each galaxy was defined as the maximum among all filters of the catalogue.}
	\label{fig:hist}
\end{figure*}

Fig. \ref{fig:hist} shows histograms for the maximum surface brightness among all filters in the range $0.1<z<8.5$ (outside this range there are too few galaxies in each bin for analysis). Twenty-five bins were used both for $z$ and for the surface brightness histogram. All bins --- both in surface brightness and in $z$ --- were computed in logarithmic coordinates: $\log SB$, $\log (1+z)$.

\begin{figure}
	\includegraphics[width=\hsize]{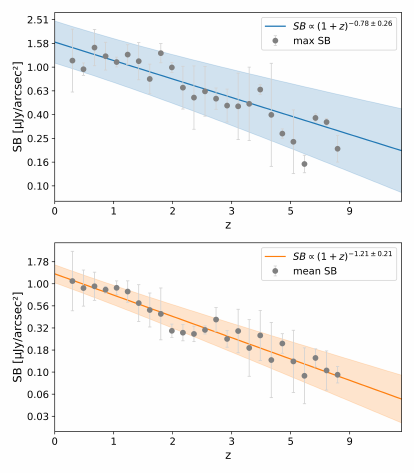}
	\caption{Surface brightness (SB) of the ``most typical'' galaxy compared with redshift (defined as the mode in the histogram within a $z$ range). Top: maximum SB; bottom: mean SB across the 16 filters of the ASTRODEEP-JWST catalogue. The determined values are shown as points. Solid lines indicate the best-fit approximation curve with bands of formal errors. The plot is shown in logarithmic axis scale.}
	\label{fig:SBhist}
\end{figure}

Next, these dependencies were approximated using a weighted least-squares method to determine the proportionality coefficient in $SB\propto (1+z)^\alpha$. Fig. \ref{fig:SBhist} shows the results for the analysis of histograms based on the maximum and the mean surface brightness (corresponding to the second and the first methods of determination described above).

The selection of a ``typical'' galaxy within bins of $z$ makes it possible to remove the uneven contribution of faint galaxies excluded by observational selection at different $z$, in accordance with the Malmquist profile, thus providing another way to estimate a reference source. In fact, Fig. \ref{fig:SBhist} represents a smoothed version of Fig. \ref{fig:SBmaxmean} and more clearly illustrates the observed dependence of the mean surface brightness of galaxies on $z$.

From the analysis of the different methods it can be seen that the surface brightness of galaxies decreases as $SB\propto(1+z)^\alpha$, where $\alpha \sim -1$. The difference between the methods lies essentially within the error limits.

\section{Analysis of Surface Brightness Evolution}

Let us consider the question of the evolution of the mean surface emission  of a galaxy as dictated within the framework of the standard cosmological model. We estimate what the mean surface emission of the galaxies under consideration would be in the two cosmological theories considered here --- the standard $\Lambda$CDM model and the reference (framework) static ``tired light'' model.

The angular size $\theta$ of a galaxy is defined as
$$\theta=\frac R{D_A},$$
where $R$ is the linear size of the galaxy and $D_A$ is the angular diameter distance. The luminosity distance is
$$D_L=\sqrt{\frac L{4\pi f_{obs}}},$$
where $L$ is the luminosity of the galaxy and $f_{obs}$ is the observed flux. The observed emission intensity $J_{obs}$ is defined as
$$J_{obs}=\frac {D^2_A}{R^2}f_{obs}.$$
For the $\Lambda$CDM model $D_L=(1+z)^2D_A$,
while for the ``tired light'' model $D_L=\sqrt{1+z}D_A$.

Thus, performing the corresponding substitutions, we obtain
$$J_{obs}^{\Lambda\mathrm{CDM}}=\dfrac {f_{obs}\frac{L}{4\pi f_{obs}}}{R^2(1+z)^4},$$
$$J_{obs}^{\mathrm{TL}}=\dfrac {f_{obs}\frac{L}{4\pi f_{obs}}}{R^2(1+z)}.$$

Taking the mean intensity of a galaxy as $J_0=\frac{L}{4\pi R^2}$, we obtain
$$J_0^{\Lambda\mathrm{CDM}}=(1+z)^4 J_{obs},$$
$$J_0^{TL}=(1+z) J_{obs}.$$

\begin{figure}
	\includegraphics[width=\hsize]{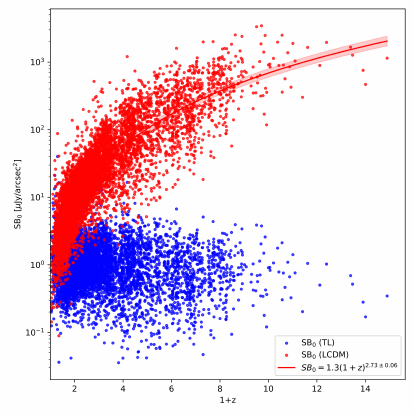}
	\caption{Galaxy surface brightness $SB_0$ in two models. The maximum surface brightness over the 16 filters of the ASTRODEEP-JWST catalogue is used. The determined values are shown as points. Solid lines show the best-fit approximation curve with bands of formal errors, illustrating the evolution of galaxy surface brightness within the $\Lambda$CDM model.}
	\label{fig:SB0}
\end{figure}

Let us recalculate the data shown in the upper panel of Fig.
\ref{fig:SBmaxmean} (the maximum surface brightness of the galaxy over the 16 filters) into $J_0$ (surface brightness $SB_0$) for the two models using these formulas. The results are shown in Fig. \ref{fig:SB0}.

In our previous work we showed that the evolution of the linear sizes of galaxies within the $\Lambda$CDM framework corresponds to the rate of spatial expansion $(1+z)$. Thus, one may expect that the mean emission intensity should at least increase with redshift as $(1+z)^{2}$ --- due to the growth of stellar density. Consequently, the observed surface brightness should follow the factor $(1+z)^{-2}$ (due to redshift and time dilation). Our results show some deviation from this dependence. However, the current accuracy, associated with the difficulty of determining bolometric surface brightness and accounting for observational selection, does not allow us to identify an unambiguous evolutionary trend, which should include both the evolution of the stellar population and the stellar density of galaxies. This requires further investigation and modelling.

Another factor influencing surface brightness is the composition of the stellar population of a galaxy, i.e., the characteristic flux of the source: it is expected that for a galaxy composed mainly of young O--B stars the surface brightness will be higher. In particular, when using the wavelength range 1500--2000 \AA{} in the rest frame, the surface brightness produced by such stars is being estimated: if the star formation intensity in a galaxy is higher, the fraction of such stars is larger and the true surface brightness will be higher. Note that for this particular variant of the analysis (Fig. \ref{fig:SBrest}) the slowest decrease of surface brightness is observed, which may indicate an mean decline in the star formation rate with age. However, this question lies beyond the scope of the present study.

At the same time, it can be seen that within the static cosmological model there is practically no evolution of the emission intensity --- just as there is no evolution of linear sizes according to our previous work in this series.

\section{Conclusion}

Compared to the cosmological test ``angular size --- redshift'', the Tolman test is a stronger one, since the dependence on redshift required by standard cosmology predicts a decrease of surface brightness proportional to the fourth, rather than the first, power of $(1+z)$. At the same time, the Tolman test is more difficult to investigate, since it requires the presence of an ideal source over the entire range of redshifts. Only the availability of a very large number of measurements over the widest possible range of $z$ makes it possible to reveal the trend characterizing the sought dependence of surface brightness on redshift. Consequently, the practical implementation of the Tolman test has become possible only with the advent of JWST observations.

As shown in this work, regardless of the method used to analyse the photometric observations of galaxies, the overall trend of the mean surface brightness shows a decrease with increasing redshift. The observed decrease of surface brightness with distance at large $z$ is close to a inverse linear dependence on $(1 + z)$.

Within the framework of standard cosmology, the observed decrease of surface brightness with distance can be explained only by a strong --- with a rate of about the third power of $(1+z)$ --- evolution of surface brightness. 
At the same time, within the static cosmological model no significant evolution of surface brightness, as well as of the linear sizes of galaxies, is detected.
\color{black}

\section*{Acknowledgements}
The ASTRODEEP-JWST catalogue used in this work is based in part on observations made with the NASA/ESA/CSA James Webb Space Telescope. The data were obtained from the Mikulski Archive for Space Telescopes at the Space Telescope Science Institute, which is operated by the Association of Universities for Research in Astronomy, Inc., under NASA contract NAS 5-03127 for JWST. These observations are associated with program JWST-ERS-1342. This catalogue is also based in part on observations made with the NASA/ESA Hubble Space Telescope obtained from the Space Telescope Science Institute, which is operated by the Association of Universities for Research in Astronomy, Inc., under NASA contract NAS 5-26555. These observations are associated with program HST-GO-17321. 

\section*{Funding}
A.~A.~Raikov carried out this work within the framework of the state assignment of the Special Astrophysical Observatory of the Russian Academy of Sciences, approved by the Ministry of Science and Higher Education of the Russian Federation.  V.~V.~Tsymbal also carried out this work within the framework of the state assignment of the Institute of Astronomy of the Russian Academy of Sciences. The other co-authors received no additional funding.

\bibliographystyle{abbrvnat}
\bibliography{Test-2}

\end{document}